\documentclass{xray_symp}
\usepackage{graphicx}
\usepackage{natbib}

\newcommand{\keV}{\mbox{keV}}
\newcommand{\eV}{\mbox{eV}}
\newcommand{\cyg}{Cyg~X-1}
\newcommand{\Msun}{\ensuremath{\rm M_\odot}}

\begin{document}

\title{RXTE Observations of Cygnus X-1}
\subtitle{Broad-Band Spectra and High-Resolution Timing}     
\author{K.~Pottschmidt\inst{1} \and J.~Wilms\inst{1} \and
M.A.~Nowak\inst{2} \and J.B.~Dove\inst{3} \and M.C.~Begelman\inst{2}
\and R.~Staubert\inst{1}}  
\institute{%
Institut f\"ur Astronomie und Astrophysik -- Astronomie,
Waldh\"auser Str. 64, D-72076 T\"ubingen, Germany
\and
JILA, University of Colorado, Boulder, CO 80309-440, U.S.A.
\and
CASA, University of Colorado, Boulder, CO 80309-389, U.S.A.}

\thesaurus{XXXXXXXX}  
\date{X}
\maketitle

\begin{abstract}  
  We present results from a 20\,ksec RXTE observation of the black hole
  candidate \cyg. We apply self-consistent accretion disk corona models to
  these hard state data and show that Comptonization in a spherical corona
  irradiated by soft photons from an exterior cold disk is able to
  successfully model the spectrum. We also present the power spectrum, the
  coherence function, and the time lags for lightcurves from four energy
  bands. By modeling the high-resolution lightcurves with stochastic linear
  state space models, we show that the rapid hard state variability of
  \cyg\ can be explained with a single timescale.
\end{abstract}

\section{Introduction}
The X-ray emission of galactic black hole candidates (BHCs) displays
different ``states'' characterized by distinct temporal and spectral
properties \citep{esin:97a,klis:95}. Except for occasional transitions to
the soft state, Cyg X-1 spends most of its time in the hard state
(Fig.~\ref{fig:asm}). It is therefore the prime candidate for evaluating
accretion models of this fundamental black hole state.

Since 1996 the Rossi X-ray Timing Explorer (RXTE) provides data from 3 to
200\,\keV\ with millisecond time-resolution. In this paper we derive a wide
variety of spectral and temporal hard state properties by applying physical
accretion disk corona models (ADCs) (Sect.~\ref{spectra}) as well as new
timing methods (Sect.~\ref{fourier} and~\ref{lssm}) to a RXTE hard state
observation of \cyg. Whereas ADCs are very successful in explaining the
observed spectrum \citep{dove:97b}, a unified model for the hard state
accretion also has to take into account the time dependence of the
emission. Some of the associated constraints that are provided by the power
spectral density, the coherence function, the time lags, and the lightcurve
modeling with stochastic processes, are discussed in sections~\ref{fourier}
and~\ref{lssm}. For further discussion see \citet{nowak:98c,nowak:98b} and
\citet{pottschmidt:98a}.

\section{Data Extraction}\label{observation}
\begin{figure}
\resizebox{\hsize}{!}{\includegraphics{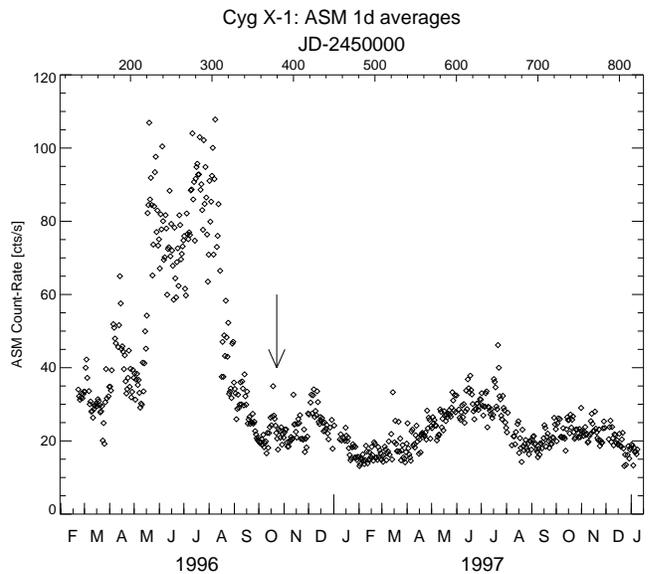}}
\caption{RXTE-ASM count-rates in the 1.5 to 12\,\keV\ band for \cyg\
  from the beginning of the ASM measurements until 1997 December. The arrow
  indicates the time of our pointed RXTE observation. The 1996 soft state
  can be clearly identified by the increased count-rate from 1996 May
  through August.\label{fig:asm}}
\end{figure}

RXTE observed \cyg\ for a total integrated source time of slightly more
than 20\,ksec, starting 1996 October~23, briefly after its transition to the
hard state (Fig.~\ref{fig:asm}). 

We analyzed data from the Proportional Counter Array (PCA) as well as from
the High Energy X-ray Timing Experiment (HEXTE). An overview of our data
screening procedures and of the RXTE instruments is given by
\citet{dove:97c} and Wilms et al. (this volume). For the spectral analysis
we used 3 to 30\,\keV\ PCA data and 20 to 200\,\keV\ HEXTE data. The temporal
analysis is based on PCA lightcurves with 1.95\,ms resolution from four
energy bands of approximately equal count rates ($\cal
O$(700--1000)\,cts\,s$^{-1}$ per band). The energy bands are defined in
Fig.~\ref{fig:psd}, the total length of the lightcurves was $\sim$18\,ksec.
More details on the PCA timing data of this observation are given by
\citet{nowak:98b}.

\begin{figure}
\resizebox{\hsize}{!}{\includegraphics{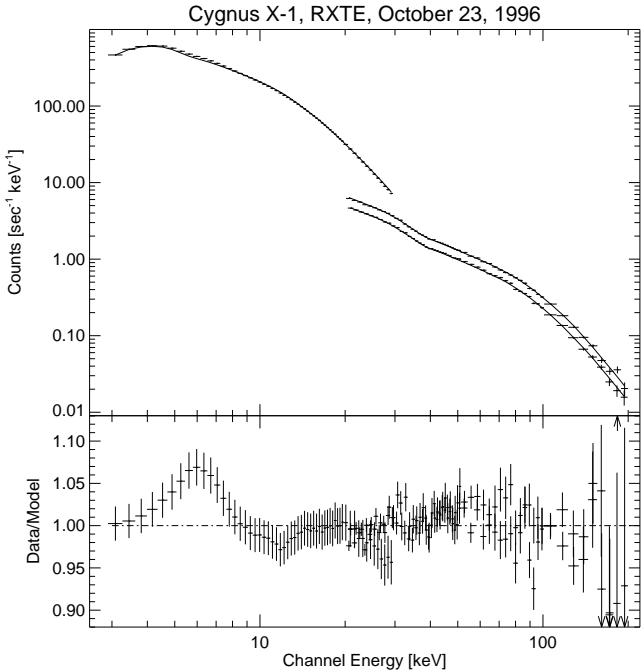}}
\caption{Fit of the sphere+disk model to the data, indicating good overall
  agreement between the ADC model and the observations ($\chi^2_{\rm
    red}=1.55$, assuming a systematic error of 2$\%$ for the PCA data). The
  corona has a temperature of $kT=(65.7\pm3.3)$\,\keV\ and an optical depth
  of $\tau=2.1\pm0.1$. The seed photons were assumed to come from an
  accretion disk external to the corona, with a $r^{-3/4}$ temperature
  profile and an inner accretion disk temperature of
  $150$\,\eV.\label{fig:spec}}
\end{figure}

\section{Spectral Modeling: Self-Consistent Thermal Accretion Disk Corona
  Models}\label{spectra} The BHC hard state spectrum, a power-law with
$\Gamma\sim1.65$, modified by an exponential cutoff with
$E_{\rm fold}\sim150$\,\keV\ is naturally explained by accretion disk
corona (ADC) models, i.e., by Comptonization of low-energy seed photons from a
cold, geometrically thin accretion disk in a hot, semi-relativistic corona
\citep{tit:94}. As we explain in more detail in an accompanying paper
\citep[Wilms et al., this volume; see also ][]{dove:97a,dove:97b}, we have
developed an ADC model in which the radiation field, the temperature and
opacity of the corona, and the reprocessing of coronal radiation in the
accretion disk are calculated self-consistently. The commonly assumed slab
geometry where the accretion disk is sandwiched between two coronae is not
self-consistent, since the coronal temperatures and opacities are not
within the the physically allowed range of combinations of these
parameters. On the other hand, the sphere+disk geometry where the seed
photons come from an accretion disk external to the inner spherical corona,
allows to fit the broad-band spectrum of \cyg\ with \emph{one}
self-consistent model.

The best-fit spere+disk spectrum for our RXTE observation is shown in
Fig.~\ref{fig:spec}. The deviations between the data and model above
10\,\keV\ are consistent with the current calibration uncertainty of the
PCA, while the deviation in the iron band between 5 and 10\,\keV\ shows the
physical simplifications in our model. This does not pose a severe problem
to the model since the deviations represent less than 1\% of the total flux
modeled. The deviation might be due to an underestimation of the covering
factor for reflection, e.g., by a possible overlap between the disk and the
corona due to the transition from the soft to the hard state two weeks
before the observation (cf.~Fig.~\ref{fig:asm}). The presence of accretion
disk flares (due to disk heating) would also increase the reprocessing
observed.  Finally, a slight increase in the Fe abundance will also produce
the stronger line seen in the data. See \citet{dove:97c} for a discussion
of these points, as well as for a presentation of spectral fits using
phenomenological ``standard'' models.

\section{Timing: Fourier Methods}\label{fourier}

\subsection{Power Spectral Density}
\begin{figure}
\resizebox{\hsize}{!}{\includegraphics{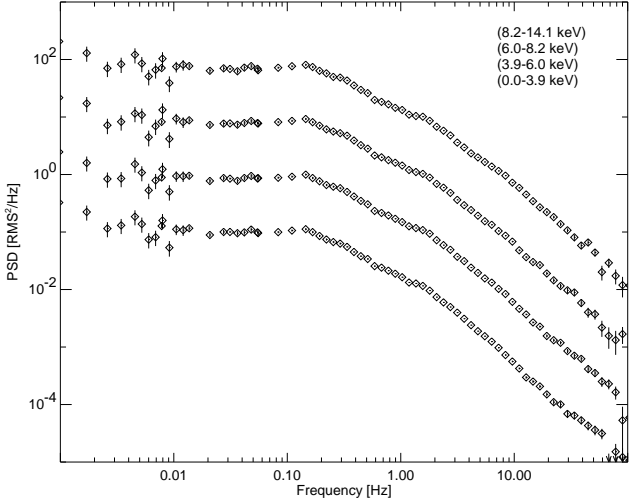}}
\caption{Energy resolved PSDs normalized according to
  \citet{miyamoto:92}. PSDs for the higher energy bands have been
  multiplied by powers of 10 for clarity. The Poisson counting-noise has
  been subtracted taking the instrumental deadtime into account
  \citep{zhangw:95a}.\label{fig:psd}}
\end{figure}

We calculated power spectral densities (PSDs) by dividing the data into
contiguous segments of uniform length and computing the
Fourier transform $S_j(f)$ of each data segment. The discrete PSD is then
given by $\langle{|{S_j}}|^2\rangle$ where the  brackets indicate an
average over data segments and logarithmic frequency bins
\citep[Fig.~\ref{fig:psd}, see also ][]{nowak:98b}. 

The PSDs show characteristic hard state behavior \citep[cf.\ 
][]{belloni:90a}. For 10$^{-3}$--100\,Hz the rms variability in all bands
is about 30\%. The PSDs are relatively flat for frequencies from
0.02--0.2\,Hz. For higher frequencies, they are roughly described by a
steepening, broken power-law with a break at $\sim$2\,Hz, and typical
slopes of $-$1.0, and $-$1.5, respectively. Note that while there are no
strong energy dependences, the high frequency slope does ``harden'' for
higher energies.

\subsection{Coherence Function}
The coherence function, $\gamma^2(f)$, is a frequency-dependent measure of
the degree of linear correlation between two concurrent time series
\citep{vaughan:97,nowak:98b}. Specifically, it gives the fraction of the
variability of one time series that can be predicted from
the other. The coherence is based on the Fourier transforms $S(f)$
and $H(f)$, calculated for the soft and the hard lightcurve, respectively:
\begin{equation}
\gamma^2(f)=\frac{|\langle{S(f)^{*}H(f)}\rangle|^2}{\langle|S(f)|^2\rangle\langle|H(f)|^2\rangle} 
\end{equation}
where the brackets indicate an average over data segments and frequency
bins and the star denotes complex conjugation.

\begin{figure}
\resizebox{\hsize}{!}{\includegraphics{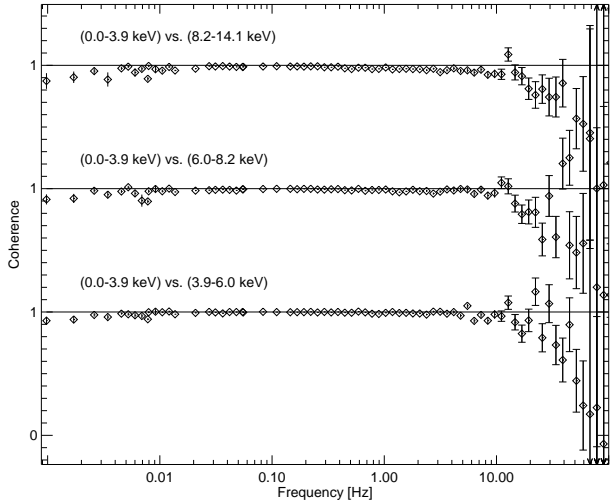}}
\caption{Coherence between the lowest energy band and the three higher
  ones, corrected for Poisson counting-noise. The noise correction and the
  error bars have been calculated according to eq.~(8) of
  \citet{vaughan:97}. Especially for high frequencies the noise-subtracted
  coherence is subject to additional systematic uncertainties
  ($<$30\% from 30 to 100\,Hz), which is why deviations to
  greater than unity can occur.\label{fig:cof}}
\end{figure}

We computed the coherence function between the lowest energy band and the
three higher ones using the same lightcurve segments and frequency
rebinning as for the PSDs.  The coherence is remarkably constant and unity
over the broad frequency range from 0.02 to 10\,Hz (Fig.~\ref{fig:cof}) --
in agreement with previous results and contrary to theoretical expectations
\citep{vaughan:97,cui:97b}. This behavior of $\gamma^2$ indicates that the
Comptonizing medium must be static on these timescales, despite the various
possibilities for non-uniform coronae \citep{nowak:98c}.

For all energy channels there is a trend for the coherence to drop below
$\sim$0.02\,Hz and above $\sim$10\,Hz with the deviation from unity
coherence generally becoming slightly greater for increasing energy
difference. The loss of coherence below 0.02\,Hz is consistent with the
viscous timescale for $R>50GM/c^{2}$, while the loss of coherence above
10\,Hz is consistent with dynamical timescale for $R<50GM/c^{2}$ ($R$ is
the distance from a 10\,\Msun\ black hole). For further discussion of the
physical implications see \citet{nowak:98c} and \citet{nowak:98b}.

\subsection{Time Lags}
The Fourier time lag is a frequency-dependent measure of the time delay
between two time series. Like the coherence it is calculated from their
Fourier transforms $S(f)$ and $H(f)$,
\begin{equation}
\tau(f)=\frac{{\rm arg}\langle{S(f)^{*}H(f)}\rangle}{2\pi{f}}
\end{equation}
$\tau$ can be either positive or negative. For our sign convention, a
positive time lag indicates that the hard lightcurve lags the soft
lightcurve.

\begin{figure}
\resizebox{\hsize}{!}{\includegraphics{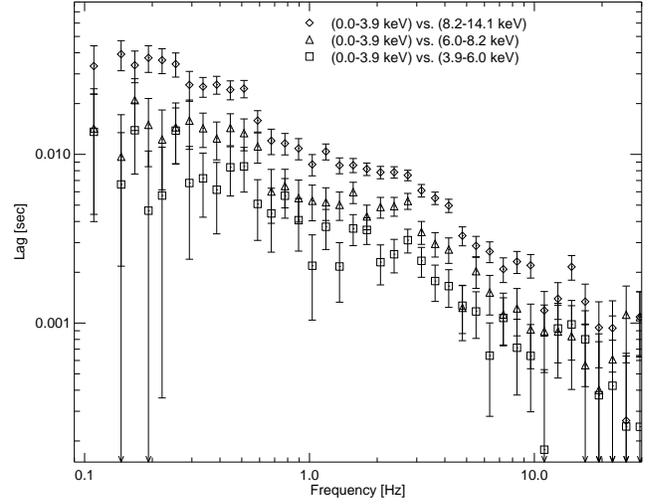}}
\caption{Time lag as a function of frequency between the lowest energy band
  and the three higher ones. The error bars are calculated
  according to eq.~(16) of \citet{nowak:98b}.\label{fig:lag}}
\end{figure}

We computed the time lags for the same energy bands, lightcurve segments,
and frequency bins as used for the coherence (Fig.~\ref{fig:lag}). Within
the frequency range that is not noise-dominated (0.1--30\,Hz), the data are
consistent with the softest energy band always leading the harder energy
bands. The lags increase significantly with the energy difference between
the bands and become larger for smaller frequencies, a result previously
found by \citet{miyamoto:92}. The time lags range from $\sim$0.002\,s to
$\sim$0.05\,s and can be very roughly described with a power law,
$\tau(f)\propto f^{-0.7}$, superposed by characteristic breaks at $\sim$0.5,
3, and 10\,Hz.

The simplest expectation in the framework of Comptonization models is that
time delays should be due to different diffusion times through the corona
and thus should be nearly independent of Fourier frequency
\citep{miyamoto:89}.  Furthermore, the overall energy dependence of the
time lags is also not compatible with current Comptonization models
\citep{nowak:98b} -- including the sphere+disk model.  It is therefore
necessary to consider more complicated accretion models \citep{nowak:98c}.
The smallest observed time lags might indicate that the Comptonizing medium
is rather small, $<30\,GM/c^2$. Consequently, if the time delays are due to
disturbances propagating through the corona, the propagation speeds are
extremly slow ($\cal O$(1--10\%\,$c$)). These values are inconsistent with
current ADAF models.

\section{Timing: The Linear State Space Model}\label{lssm}
An alternative method to the timing analysis in the frequency domain is to
examine the X-ray variability directly in the time domain by modeling the
observed lightcurves with linear state space models (LSSMs)
\citep{koenig:97a}. These models describe the dynamics of a lightcurve
$y(t)$ with a stochastic, autoregressive (AR) process $x(t)$ of order $p$
\citep{scargle:81}:
\begin{equation}
  x(t)=\left(\sum_{j=1}^p{a_j\cdot{x(t-j)}}\right)+\epsilon(t)
\end{equation}
where $\epsilon(t)$ is a Gaussian white noise processes with zero mean and
variance $\sigma_{\epsilon}^2$.  The dynamical parameters, $a_j$, are
closely related to the temporal parameters characterizing the system. An
AR[p=1] process, e.g., is described by a single relaxation timescale
$\tau=-1/\ln{(a_1)}$.

LSSMs also take the observational noise into account. For an AR[1]
process 
\begin{equation}
  y(t)=c\cdot x(t)+\eta(t)
\end{equation}
where $\eta(t)$ is the observational noise, again given by a Gaussian white
noise processes with zero mean and a variance of $\sigma_{\eta}^2$, and
where $c$ is a constant.

We fitted LSSMs of different order to 842 contiguous lightcurve segments
(0.0--14.1\,\keV) with a length of 16\,s each. To evaluate the goodness of
the fits, a Kolmogorov-Smirnov test for white noise residuals was
performed. We find that the lightcurves can be modeled with a LSSM[1]
process. This is consistent with our results for modeling EXOSAT
observations of \cyg \citep{pottschmidt:98a} and in analogy to the latter
work we derive a relaxation timescale $\tau=(0.13\pm0.03)$\,s for the RXTE
lightcurves. In contrast to the usually applied shot noise model fits in
the frequency domain \citep{lochner:91}, the LSSMs of first order do not
require pre-defined shot forms and are able to describe the rapid hard
state variability of \cyg\ with \emph{one} temporal parameter.

The derived relaxator can in principle be used for scaling time dependent
ADC models, by fitting LSSMs to theoretical ``ADC lightcurves''. First
estimates indicate much larger coronal radii than implied by the shortest
time lags ($\sim600\,GM/c^2$).  For a better understanding of spectral und
temporal hard state properties in one unified accretion scenario, we plan
to apply the models and methods presented here to further RXTE observations
of \cyg\ and other BHCs.

\acknowledgements This work has been financed by NSF grants AST91-20599,
AST95-29175, INT95-13899, NASA Grant NAG5-2026, NAG5-3225, NAGS-3310, DARA
grant 50\,OR\,92054, and by a travel grant to J.W.\ and K.P.\ from the
DAAD.

\end{document}